\documentclass[aps,prl,twocolumn,amsmath,amssymb,superscriptaddress]{revtex4-1}
\usepackage{graphicx}
\usepackage{dcolumn} 
\usepackage{graphics}

\begin{document}
\title{Gate-Sensing Charge Pockets in the Semiconductor Qubit Environment}
\author{X. G. Croot}
\affiliation{ARC Centre of Excellence for Engineered Quantum Systems, School of Physics, The University of Sydney, Sydney, NSW 2006, Australia.} 
\author{S. J. Pauka}
\affiliation{ARC Centre of Excellence for Engineered Quantum Systems, School of Physics, The University of Sydney, Sydney, NSW 2006, Australia.} 
\author{H. Lu}
\affiliation{Materials Department, University of California, Santa Barbara, California 93106, USA.}
\author{A. C. Gossard}
\affiliation{Materials Department, University of California, Santa Barbara, California 93106, USA.}
\author{J. D. Watson}
\affiliation{Department of Physics and Astronomy, Purdue University, West Lafayette, IN 47907, USA.}
\affiliation{Birck Nanotechnology Center, Purdue University, West Lafayette, IN 47907, USA.}
\author{G. C. Gardner}
\affiliation{Station Q Purdue, Purdue University, West Lafayette, IN 47907, USA.}
\affiliation{Birck Nanotechnology Center, Purdue University, West Lafayette, IN 47907, USA.}
\author{S. Fallahi}
\affiliation{Department of Physics and Astronomy, Purdue University, West Lafayette, IN 47907, USA.}
\affiliation{Birck Nanotechnology Center, Purdue University, West Lafayette, IN 47907, USA.}
\author{M. J. Manfra}
\affiliation{Station Q Purdue, Purdue University, West Lafayette, IN 47907, USA.}
\affiliation{Department of Physics and Astronomy, Purdue University, West Lafayette, IN 47907, USA.}
 \affiliation{School of Materials Engineering and School of Electrical and Computer Engineering, Purdue University, West Lafayette, IN 47907, USA.}
 \affiliation{Birck Nanotechnology Center, Purdue University, West Lafayette, IN 47907, USA.}
\author{D. J. Reilly$^\dagger$}
\affiliation{ARC Centre of Excellence for Engineered Quantum Systems, School of Physics, The University of Sydney, Sydney, NSW 2006, Australia.} 
\affiliation{Microsoft Corporation, Station Q Sydney, The University of Sydney, Sydney, NSW 2006, Australia.} 

\begin{abstract}
We report the use of dispersive gate sensing (DGS) as a means of probing the charge environment of heterostructure-based qubit devices. The DGS technique, which detects small shifts in the quantum capacitance associated with single-electron tunnel events, is shown to be sensitive to pockets of charge in the potential-landscape likely under, and surrounding, the surface gates that define qubits and their readout sensors. Configuring a quantum point contact (QPC) as a localised emitter, we show how these charge pockets are activated by the relaxation of electrons tunnelling through a barrier. The presence of charge pockets creates uncontrolled offsets in gate-bias and their thermal activation by on-chip tunnel currents suggests further sources of charge-noise that lead to decoherence in semiconductor qubits.
\end{abstract}
\maketitle

\begin{figure}
\includegraphics[scale=0.14]{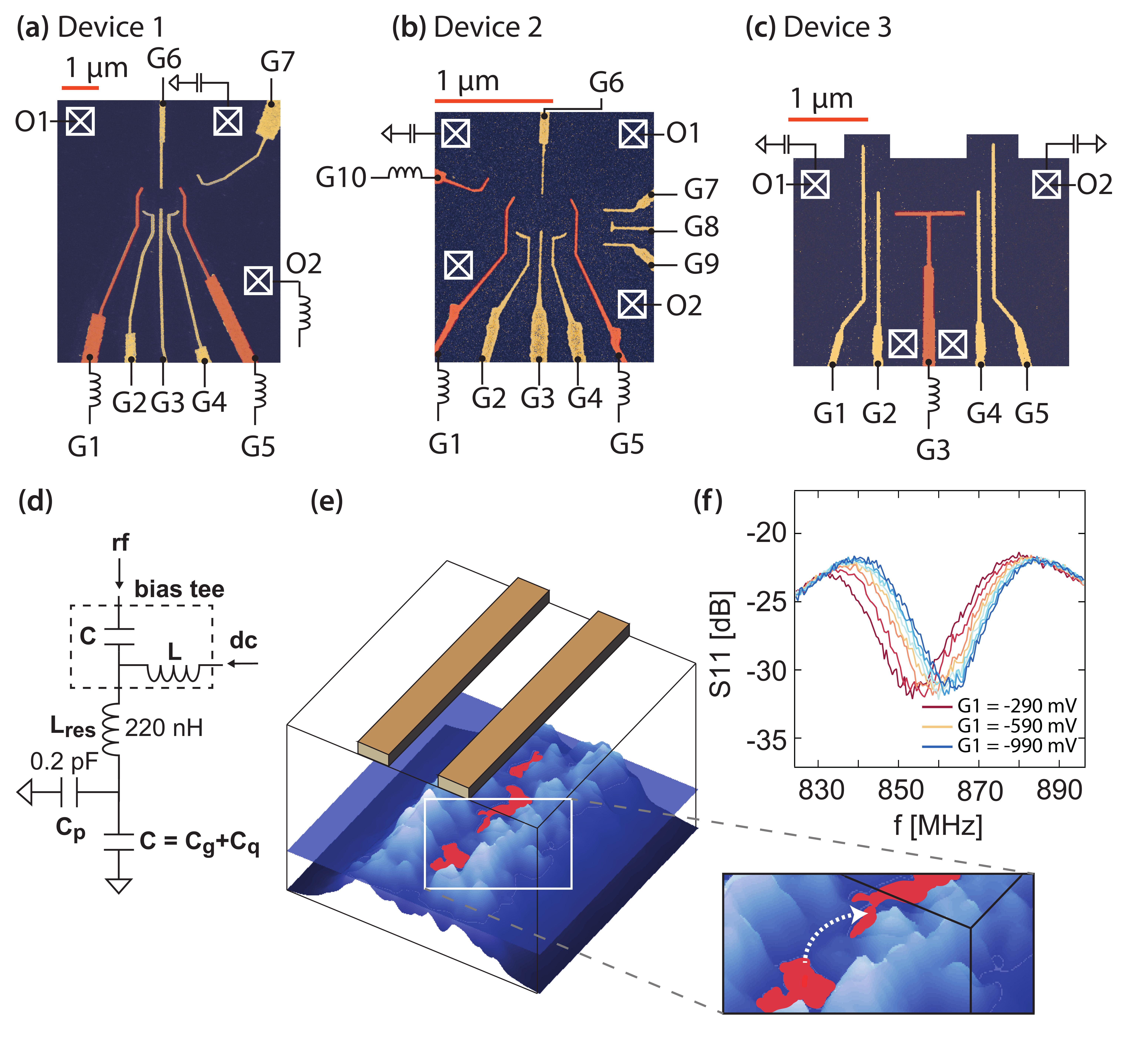}
\caption{\label{} (a-c) False-coloured micrographs of the three devices examined. Each device is fabricated from a unique GaAs/AlGaAs heterostructure with mobilities of 3.9, 0.44, and 2.4 $\times 10^{6}$ cm$^2/Vs$, and densities 1.2, 2.4, and 1.5 $\times 10^{11}$ cm$^{-2}$, and 2DEG depths of 91 nm, 110 nm, and 91 nm for device 1, 2, and 3 respectively. White crossed boxes indicate ohmic contacts. Resonators, required for dispersive gate sensing, are indicated by the inductor symbols, with full circuit shown in (d), including parasitic capacitance $C_p$ and classical gate capacitance $C_g$. (e) Cartoon illustrating charge pockets that give rise to closely spaced Coulomb blockade oscillations in the DGS readout signal. (f) shows the frequency response of a typical resonator (attached to gate G1 of device 1) as the gate is biased from -290 mV to -990 mV.}
\end{figure}

The pristine, two-dimensional (2D) interface created in the epitaxial growth of a semiconductor heterostructure underpins much of modern mesoscopic physics and serves as a foundation for hosting quantum information, encoded in the spin-state of electrons \cite{Hanson:2007eg} or parity of Majorana zero-modes \cite{Majorana2D}. Despite their near-perfect crystallinity \cite{Gardner_Fallahi_Watson_Manfra_2016}, hetero-interfaces still contain unaccounted sources of charge noise that limit the performance of qubit devices \cite{PhysRevLett.110.146804,Zoo_2Q}. Even the presence of static, but unintentional, charges in the material is problematic, since each qubit then requires uniquely-tuned gate voltages to compensate the offset-charge from the disorder potential \cite{Meno_scale}. For semiconductor quantum systems, identifying and suppressing all sources of charge-offset and noise is essential if qubits are to be scaled-up into dense arrays under autonomous control.

Directly probing trapped-charge and inhomogeneities in the potential-landscape has long-posed a challenge for standard transport measurements, requiring alternative methods such as scanned-probe techniques \cite{Finkelstein90} that can, for instance, image electron-hole puddles \cite{PhysRevB.84.115442} at the surface of materials such as graphene \cite{Yacoby_scanned}. Puddles of charge have also been detected by measuring velocity-shifts in the propagation of surface acoustic waves in low-density 2D systems \cite{Tracy} or via the use of capacitive-bridges \cite{Ashoori_science} and local electrometers \cite{Ilani1354}.

In this Letter, we exploit the recently-pioneered technique of dispersive gate sensing (DGS) \cite{Colless_PRL} to probe the 2D potential-landscape of qubit devices in search of unaccounted sources of charge-offset and noise that leads to qubit dephasing \cite{PhysRevLett.110.146804,Yacoby2qubit}. By sensing small shifts in the quantum capacitance of a surface gate, DGS can directly detect weakly-bound charge that accumulates in pockets associated with local minima in the interface potential. The presence of trapped charge manifests as a rapidly-oscillating signal with gate voltage in the dispersive response of the sensor, consistent with Coulomb blockade from large, shallow quantum dots that are inadvertently formed by inhomogeneities in the potential at low density \cite{PhysRevB.41.7929}. 

Unlike highly-localised charge-sensing measurements based on quantum point contacts (QPCs) \cite{Reilly:2007ig} or single electron transistors (SETs) \cite{Devoret_schoel}, the DGS technique is able to probe charge pockets that accumulate under, or surrounding the entire perimeter of a gate electrode. We further investigate how such pockets are activated by the emission of phonons associated with transport through a proximal QPC tunnel barrier. Beyond enabling an estimate of the pocket charging energy, these measurements show how the potential-landscape is perturbed by the routine electrical readout and operation of the qubits, likely contributing to the non-markovian noise of the semiconductor qubit environment.

Turning to the details of our experiments, Fig. 1(a) shows three separate GaAs/AlGaAs devices with distinct gate configurations defined using electron beam lithography and TiAu metalization. The growth of the heterostructure material spans separate molecular beam epitaxy machines, and each device has been examined over multiple cooldowns and in different dilution refrigerators. The devices are also different in terms of  their carrier density, mobility, and depth of the 2DEG from the surface (for details see the caption of Fig. 1).  In the case of device 3, the TiAu gate electrodes are separated from the GaAs by an 8 nm insulating barrier of hafnium oxide (HfO), deposited using atomic layer deposition. Devices 1 and 3 were cooled with positive bias \cite{PhysRevB.72.115331}. 

\begin{figure}
\includegraphics[scale=0.27, trim = 40 40 0 0]{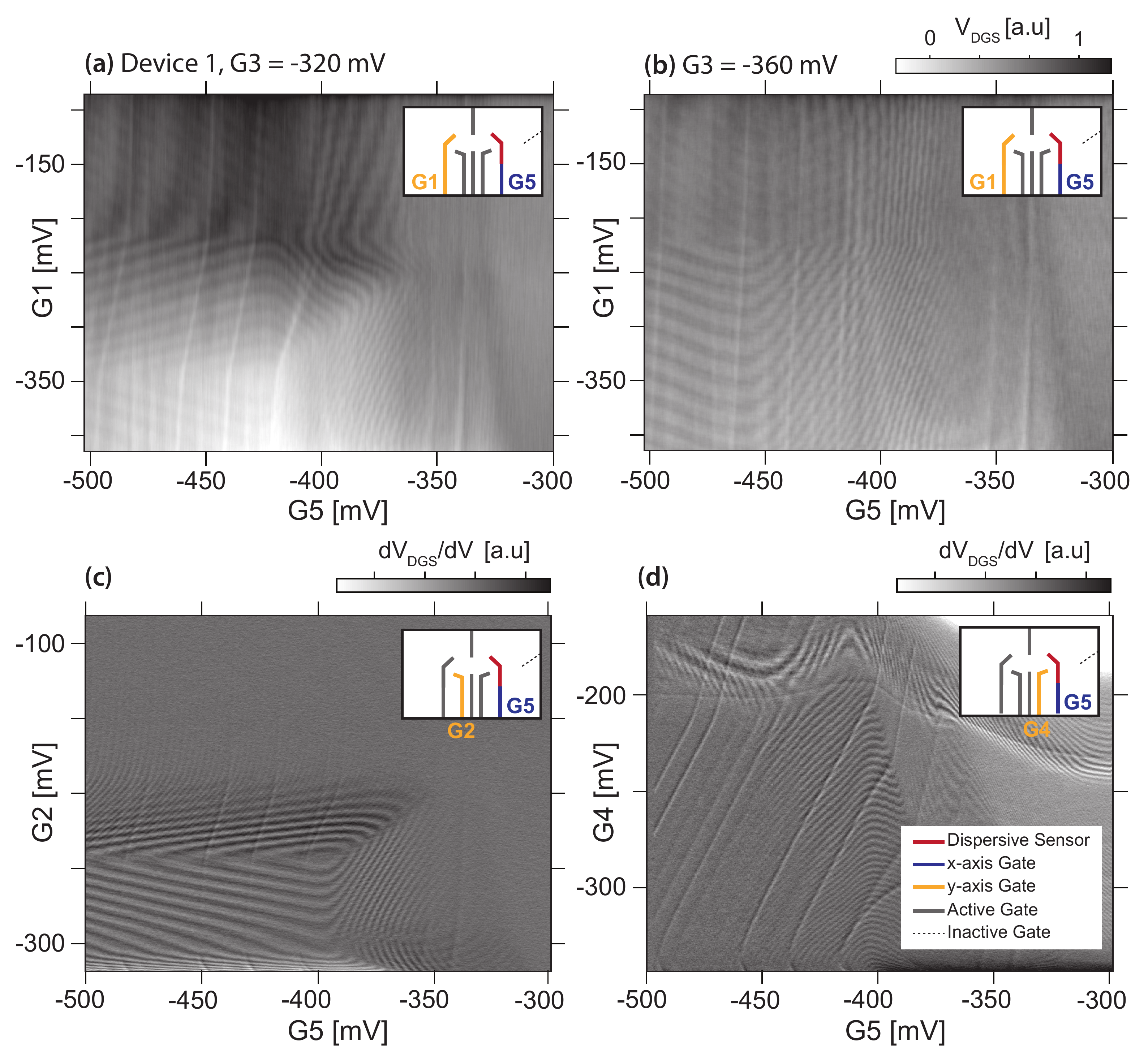}
\caption{\label{} (a) and (b) Complex, oscillatory pattern in the DGS response for device-1, as a function of gates G1 and G5, adjusting G3 by 40 mV between (a) and (b). This pattern does not resemble a typical DGS signal for a quantum dot.  (c) and (d) Derivative of $V_{DGS}$ with respect to gate bias, now as a function of G2 and G4. Active gates are held at constant potential and inactive gates at zero (see legend in (d).}
\end{figure}

Gates coloured orange in Fig. 1(a-c) are wired-bonded to radio-frequency $LC$ tank circuits to enable dispersive readout using rf-reflectometry \cite{Colless_PRL, Hornibrook_APL}. In this configuration, the capacitive component of the resonator comprises both parasitic $C_p$ and quantum $C_q$ contributions, as shown in Fig. 1(d). A typical response of a resonator with frequency, shown in Fig. 1(f), depends strongly on the gate voltage which alters the quantum capacitance in the region of the gate electrode. With all other gates held at 0 mV, stepping gate G1 from low bias to a bias that fully depletes the 2DEG underneath the gate, shifts the resonant frequency (or phase response of the resonator) as the reactance of the circuit changes. For subsequent figures, this phase response is detected by mixing-down the reflected rf-carrier to baseband, yielding a voltage $V_{DGS}$ proportional to the change in resonator reactance. 

The phenomenology of our measurements is captured in the cartoon of Fig. 1(e), which depicts a surface-gate biased to partially deplete the electron gas. As the electron density is reduced, the homogeneous 2DEG breaks-up into shallow puddles of charge, separated by tunnel barriers. The spatial distribution of such puddles is well-understood \cite{PhysRevB.41.7929,Ilani1354} to reflect the configuration of partially-ionized silicon donor sites in the AlGaAs, surface charge arrangement, and crystal disorder at the heterostructure interface. 

As the gate bias is varied, the presence of these disorder-induced charge pockets leads to tunnelling transitions which can be detected with the dispersive gate sensing technique. Figure 2 presents representative data sets in which the response of the gate-sensor exhibits oscillatory patterns under various configurations of the dc gate bias (see caption for detailed explanation). Although the particular gate-pattern was designed to produce quantum dot qubits with tunnel-coupling to the source-drain reservoirs, for the present study we intentionally do not bias the gates to values that would typically form a quantum dot. Focusing on device-1, Fig. 2(a) shows the response of the gate-sensor $V_{DGS}$ as a function of the  gates G1 and G5, with the other gates held at constant bias. In this regime the sensor response exhibits a complex pattern of lines that do not resemble the signal expected for an intentional quantum dot \cite{Colless_PRL}. Instead, the pattern of lines changes amplitude, period, and slope with gate-bias. A small variation in the bias of G3 dramatically alters the pattern [see Fig. 2(b)], and demonstrates that the signal originates from the electron gas.

We acquire and average data-sets using standard reflectometry techniques \cite{Reilly:2007ig,Colless_PRL}. To make it easier to see the fine details in these complex patterns, we plot the derivative of the sensing signal with respect to gate voltage, as shown in Fig. 2(c) and 2(d), now as function of G2 and G5. Interpreting the lines in gate-space as charge transitions between charge pockets, we note that the slope of the lines with respect to the gate-bias cannot correspond to the formation of the usual quantum dot between the gates. Rather, these transitions presumably arise from charge motion directly under and surrounding the gate electrodes, but sufficiently close to the central region of the device to be sensitive to small variations in any gate bias. Further evidence that this is the case is given by the frequency of the oscillations with respect to gate voltage, indicating that the capacitance between the gate and charge pocket is roughly a factor of 5 larger than the gate-capacitance typically observed for intentional quantum dots \cite{Hanson:2007eg}. Although not completely understood, we suggest that the curvature and changing slope of the lines relates to the complicated shape of the charge pocket and its response to strong fringing-fields from the gates, as well as the distance, orientation, and direction of tunnelling, relative to the gate-sensor \cite{Ilani1354}.

\begin{figure}
\includegraphics[scale=0.28, trim = 40 40 0 0]{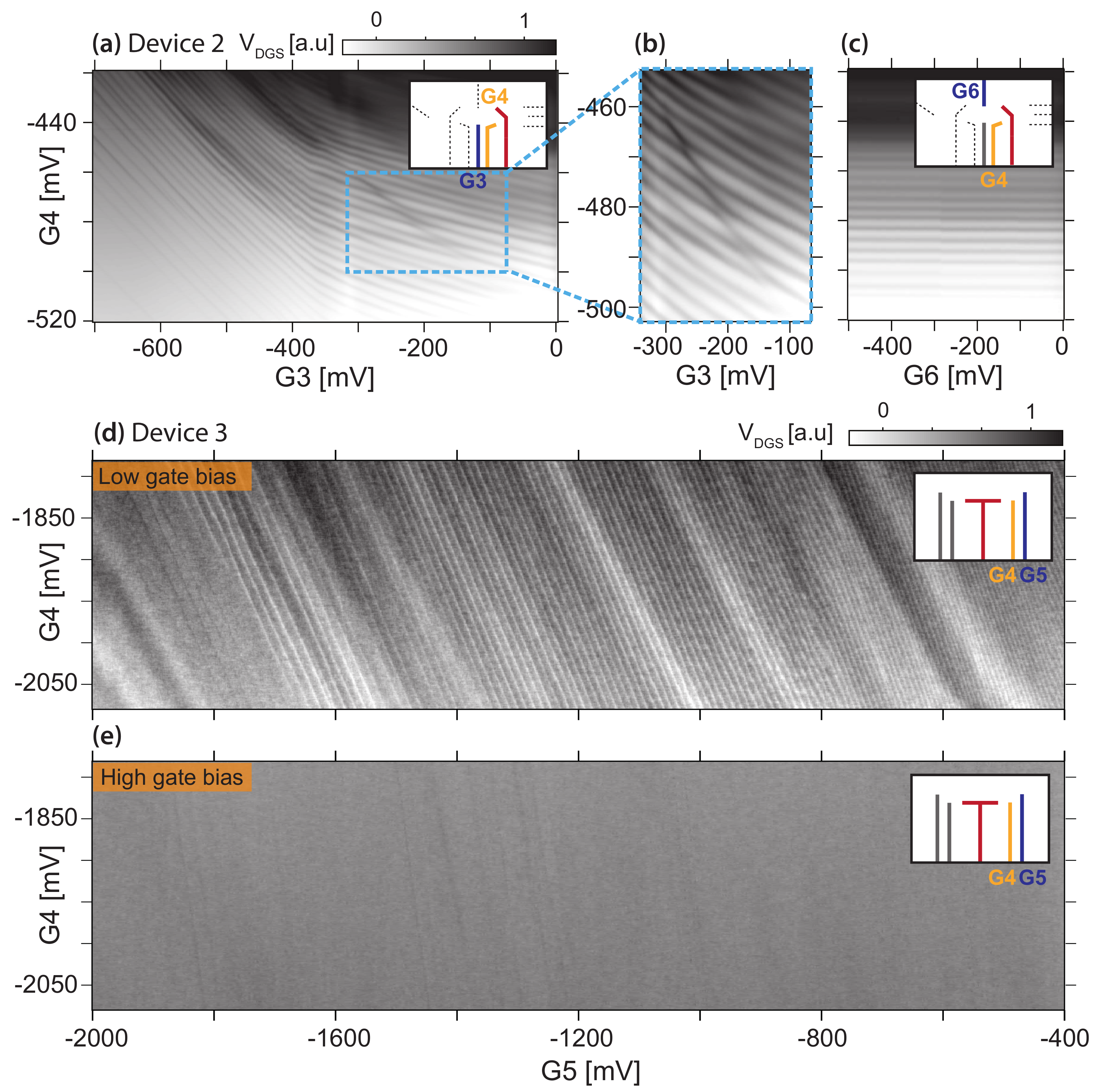}
\caption{\label{} (a) DGS response for device-2, as a function of bias on G3 and G4, with G1, G2 and G6 held at 0 V to ensure that a quantum dot is not intentionally formed. (b) Close inspection of (a) reveals an avoided crossing. (c) DGS signal as a function of G6 and G4. The response is insensitive to the bias applied to G6. (d) For device-3, a comparison between the DGS response when all gate are biased to low negative gate voltages, and (e), at high negative gate voltages, where the oscillations are suppressed.}
\end{figure}

In what follows, we pursue this charge-pocket interpretation as an explanation for the complex patterns observed with gate sensing, gathering further evidence from measurements on additional devices.  Switching to device-2, for instance, we again observe oscillatory structure in the gate sensor response, as shown in Fig. 3(a). In an effort to further pinpoint the source of this signal we limit the gate bias to three gates, holding the other gates at zero to ensure that a quantum dot cannot be formed in the central region.  Never-the-less, even with 3 gates, close inspection of the data in Fig. 3(a) [see zoomed region in Fig. 3(b)] reveals the presence of avoided-crossings in the DGS signal and provides additional evidence that we are detecting interacting charge pockets in the potential landscape, rather than the usual, gate-defined quantum dots. Of interest, applying a bias to the upper gate, $G6$, is seen to have no effect on the data, as shown in Fig. 3(c).

The strongest evidence that the oscillatory patterns are associated with charge pockets in the potential landscape is presented in Fig. 3(d) and (e), with data taken now on yet a third device, (device-3). Here we compare the gate-sensor response, first with all other gates at low bias [Fig. 3(d)], and then with all other gates set to highly negative voltages, well past the typical bias required to deplete the electron gas. The effect of this high gate-bias regime, which expels trapped charge under the gates and in the surrounding perimeter of the electron gas, is to suppress nearly all traces of the oscillatory response in the gate sensor. Finally, we note that in the case of device-3, the surface gates are insulated from the GaAs by a thin layer of HfO. Despite the presence of the HfO, the oscillatory structure in the readout persists (at low gate voltage), discounting explanations based on surface charge-states or gate-leakage, which would otherwise be modified by the addition of an insulating layer.

\begin{figure}
\includegraphics[scale=0.38]{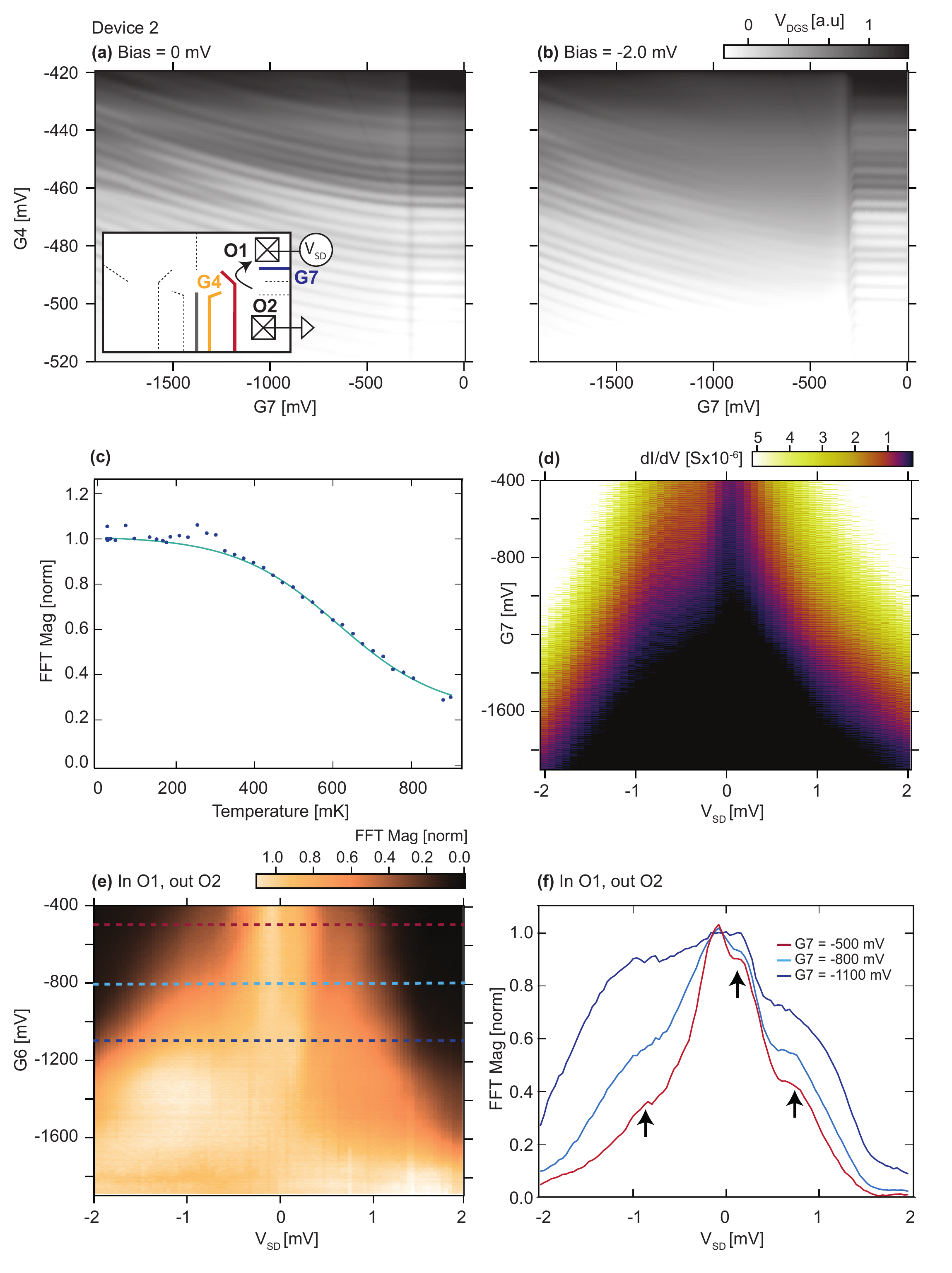}
\caption{\label{} Dependence of the DGS oscillations on nearby QPC transport for Device-2. (a) DGS response as a function of G7 and G4, without a source-drain bias across the QPC. (b) DGS oscillations with QPC bias of $V_{SD}$ = - 2.0 mV. 
The vertical line feature near - 300 mV is associated with the QPC gates depleting the electron gas. (c) Temperature dependence of the DGS oscillations, as quantified by their Fourier amplitude, normalised with respect to their amplitude at base temperature of $\sim$ 20 mK (see supplemental materials for details of the FFT analysis). Line is a guide to the eye. (d) Differential conductance of the QPC as a function of source-drain bias $V_{SD}$. (e) Amplitude of the DGS oscillations, quantified as the magnitude of their Fourier component, as a function of gate bias G7 and $V_{SD}$. (f) Horizontal 1D line-cuts of the data in (e) at positions indicated by dashed lines.}
\end{figure}

The formation of charge pockets under and between the gate-electrodes has potential implications for understanding the qubit noise environment, as well as explaining the spread in gate pinch-off voltages that stem from uncontrolled off-set charges. In this context it is worth noting that heterostructure qubits are typically defined using gate biases that produce only a partial depletion of the 2DEG, almost guaranteeing the formation of charge pockets. We next address whether these pockets can be disturbed by on-chip operations, such as qubit control or readout when using a proximal QPC or SET charge detector.

Taking device-2 as an example, we bias gates G7 and G5 to configure a QPC readout sensor, partitioning the source and drain reservoirs that connect ohmic contacts O1 and O2. At low gate bias, with the QPC open and fully transmitting, the presence of a current between O1 and O2 has little effect on the oscillations in the DGS signal. When the QPC is partially closed however, the presence of a source-drain bias, $V_{SD}$, leads to a suppression in the oscillatory signal from the gate sensor, as indicated by comparing Figs. 4(a) ($V_{SD}$ = 0) to  Fig. 4(b) ($V_{SD}$ = -2 mV). The oscillations are restored when the QPC is fully pinched-off. This sensitivity to the partial transmission of the QPC suggests that the charge pockets are activated by the emission of phonons with flux proportional to the QPC partition current, and energy proportional to the potential difference between source and drain. Raising the temperature of the cryostat above $T$ $\sim$ 200 mK is also found to strongly suppress the amplitude of oscillations in the DGS signal, as shown in Fig. 4(c) [details in supplemental material], presumably as the thermal energy becomes comparable to the charging energy of the pocket. Consistent with a large gate-capacitance that produces rapid oscillations in the DGS signal, we extract a charging energy from the temperature data that is of order a few 10s of $\mu$eV, an order-of-magnitude smaller than the typical charging energies measured for intentional, gate-defined quantum dots used as a qubits \cite{Hanson:2007eg}.

Returning to the effect of the QPC sensor on the pockets, we make a more detailed examination by first measuring the QPC differential conductance, as shown in Fig. 4(d). As is the case when charge-sensing, the QPC is very close to pinch-off, with an appreciable conductance only appearing at low gate bias and high $V_{SD}$. Next, we quantify the amplitude of the DGS oscillations by taking their fast Fourier transform (FFT) over a window of data as a function of $V_{SD}$ and gate-bias, G7, as shown on the intensity axis in Fig. 4(e) and as 1D line-cuts in Fig. 4(f) [see supplemental material for details of FFT analysis]. In this way we are making use of the DGS signal from the pockets to locally-probe the back-action of the QPC, arising from the tunnelling of electrons from source to drain \cite{PhysRevLett.102.186801,Granger}. These electrons emit phonons as they relax and thermalize in the reservoir, which then quench the small charging energy of the charge pockets. We draw attention to the appearance of step-like features that occur in the FFT-data [shown with arrows in Fig. 4(f)]. The extent to which these step-like features arise from the one-dimensional sub-bands of the QPC, or the discrete energy spectrum of the charge pockets, is an open question.
 
Having now made the case for charge pockets as the explanation for the complex oscillatory signals observed with dispersive gate sensing, we turn to further discuss their origin. In this regard, it is worth noting that such oscillatory patterns in gate-space are very rarely observed using QPC or SET charge sensors. On the other hand, we find they can always be found using DGS, even across different heterostructures (with varying mobility and density) and distinct gate patterns. Considering that the oscillations, detected by the gate incorporating the resonator, can easily be modulated by small voltages on neighbouring gates, we conclude that the location of the pockets is within a few microns from the tip of the gates. Given their small charging energy, it is likely that such pockets correspond to shallow, micron-scale, quantum dots that form directly under the gates as the electron gas is partially depleted. In such a scenario, screening from the gate metal presumably makes them difficult to detect using standard charge sensing, in contrast to DGS where the pockets contribute directly to the quantum capacitance of the resonator. 

Finally, we draw attention to the fact that these shallow pockets are easily perturbed by proximal QPC transport, and considering that qubits are operated by rf gate-pulses or microwaves, it is likely that their presence can lead to charge fluctuations in the qubit environment. The extent to which these pockets can be alleviated via the use of bi-polar, induced electron device structures \cite{PhysRevLett.89.246801,PhysRevApplied.6.054013} is an open direction for mitigating noise and offset charges in semiconductor qubits. \\

$\dagger$ Corresponding author, email: david.reilly@sydney.edu.au 

This research was supported by Microsoft Station-Q, the US Army Research Office grant W911NF-12-1-0354, the Australian Research Council Centre of Excellence Scheme (Grant No. EQuS CE110001013). We thank A.C. Mahoney for technical assistance and discussions, and acknowledge the contributions of J.I. Colless, for the fabrication of device-2, and J.M. Hornibrook, for the development of the readout multiplexing chips.

\end{document}